# BILBO-friendly Hybrid BIST Architecture with Asymmetric Polynomial Reseeding

Elahe Sadredini[1,2], Mohammadreza Najafi[2], Mahmood fathy[1] and Zainalabedin Navabi[2]
el_sadredini@comp.iust.ac.ir, mr.najafi@ut.ac.ir, mahfathy@iust.ac.ir, navabi@cad.ut.ac.ir

Computer Engineering Department[1]
Iran University of Science and Technology
16846-13114 Tehran, Iran

Electrical and Computer Engineering Department[2]
Faculty of Engineering, Campus #2
University of Tehran, 14399 Tehran, Iran

*Abstract*—**By advances in technology, integrated circuits have come to include more functionality and more complexity in a single chip. Although methods of testing have improved, but the increase in complexity of circuits, keeps testing a challenging problem. Two important challenges in testing of digital circuits are test time and accessing the circuit under test (CUT) for testing. These challenges become even more important in complex system on chip (SoC) zone. This paper presents a multi-stage test strategy to be implemented on a BIST architecture for reducing test time of a simple core as solution for more global application of SoC testing strategy. This strategy implements its test pattern generation and output response analyzer in a BILBO architecture. The proposed method benefits from an irregular polynomial BILBO (IP-BILBO) structure to improve its test results. Experimental results on ISCAS-89 benchmark circuits show an average of 35% improvement in test time in proportion to pervious work.**

***Keywords-component; BIST; DFT; BILBO; hybrid; reconfigurable; SoC testing***

## I. Introduction

As a result of advances in technology of integrated digital circuits both in size and dimension, more complex and dense circuits have evolved. Although test and testing strategies have also moderately improved, new challenges introduced by this progress necessitate more improvements in this area. The issue of porting test data to specific cores for complex systems (e.g., SoCs) has become more important due to such improvements. As a system gets larger and more complex, the cost of providing test data for its internal components increases. Build-in self-test (BIST) methods [1, 5] are proposed to mitigate the decrease in efficiency and applicability of off-chip testers. In these methods, a system includes extra components to facilitate the testing process. Multi input signature registers (MISR), pseudo random pattern generators (PRPG), and built-in logic block observer (BILBO) architectures [1, 6, 7] are examples for these units. Random test socket (RTS) is one of the most common BIST architectures that uses chained internal registers for applying test patterns and analyzing test results [7]. Utilizing chained registers increases accessibility and observability of sequential circuits under test (CUT). Thereby as expected, the achieved fault coverage in BIST methods with chained registers is relatively high.

BIST methods usually use three sources for providing their test data. The first source is using pseudo-random test generation structures like linear feedback shift register (LFSR). Low cost test vectors could be achieved from this source, but as a drawback, the obtained fault coverage is relatively low as compared with deterministic test vectors. The second source is deterministic test data stored in internal memories units of the CUT. Although using this source could lead to high coverage, the cost of dedicating expensive internal memory to the test vectors degrades efficiency of using this source. As the third source, a BIST circuit can provide its test data from an external source. This also has the disadvantage of employing expensive communication infrastructure for the test data which can negatively affect the performance of the system.

Many works have been done for improving LFSR based BIST architectures [12-14], LFSR reconfiguration [15], and optimized reseeding [16-18] to get high fault coverage and less test application time. In order to achieve high fault coverage alongside with low test application time, some works [8-11] have attempted to use hybrid BIST methods that include scan based approaches. The work in [19] has proposed a hybrid BIST method which makes use of both internally generated pseudo random test data and test data from external sources.

In this work a complete hybrid BIST architecture is proposed which employs a combination of tests from a multistage pseudo random test pattern generation method, and an external deterministic test data set. The proposed hybrid BIST makes an effort to reduce test time by decreasing the number of deterministic test vectors without affecting the overall fault coverage.

The rest of this paper is organized as follows: Section II describes proposed BIST architecture. Section III presents a methodology to apply proposed test method. In Section IV, the evaluation method is expressed. Results are illustrated in Section V and finally conclusions are drawn in Section VI.

## II. Proposed Architecture

Build-in self-test structures facilitate the testing process by integrating some or complete parts of testing components for test generation, test application, and result observation. As a result, in BIST-based architectures the communication cost of test vectors and cost of external automatic test pattern generation (ATPG) reduces significantly. Because these methods lessen the use of communication resources, they can be used as online testing methods. This is because some components of system can do their normal jobs while others are being tested. This is specially the case in multi processor systems in which system components have relatively independent tasks.

RTS is a BIST structure that uses internal registers as a scan chain. This architecture gives CUT observability and controllability just like a combinational circuit, but as a drawback it needs too many test cycles for achieving a relatively appropriate coverage.

Our proposed architecture benefits from chained registers, but instead of large test time of RTS, requires fewer test cycles. This is achieved by a new two-phase pseudo random test pattern generator unit. This component applies test data in a parallel fashion instead of serial shifting method used in RTS. An overall schematic of the proposed hybrid BIST method is shown in Fig. 1.

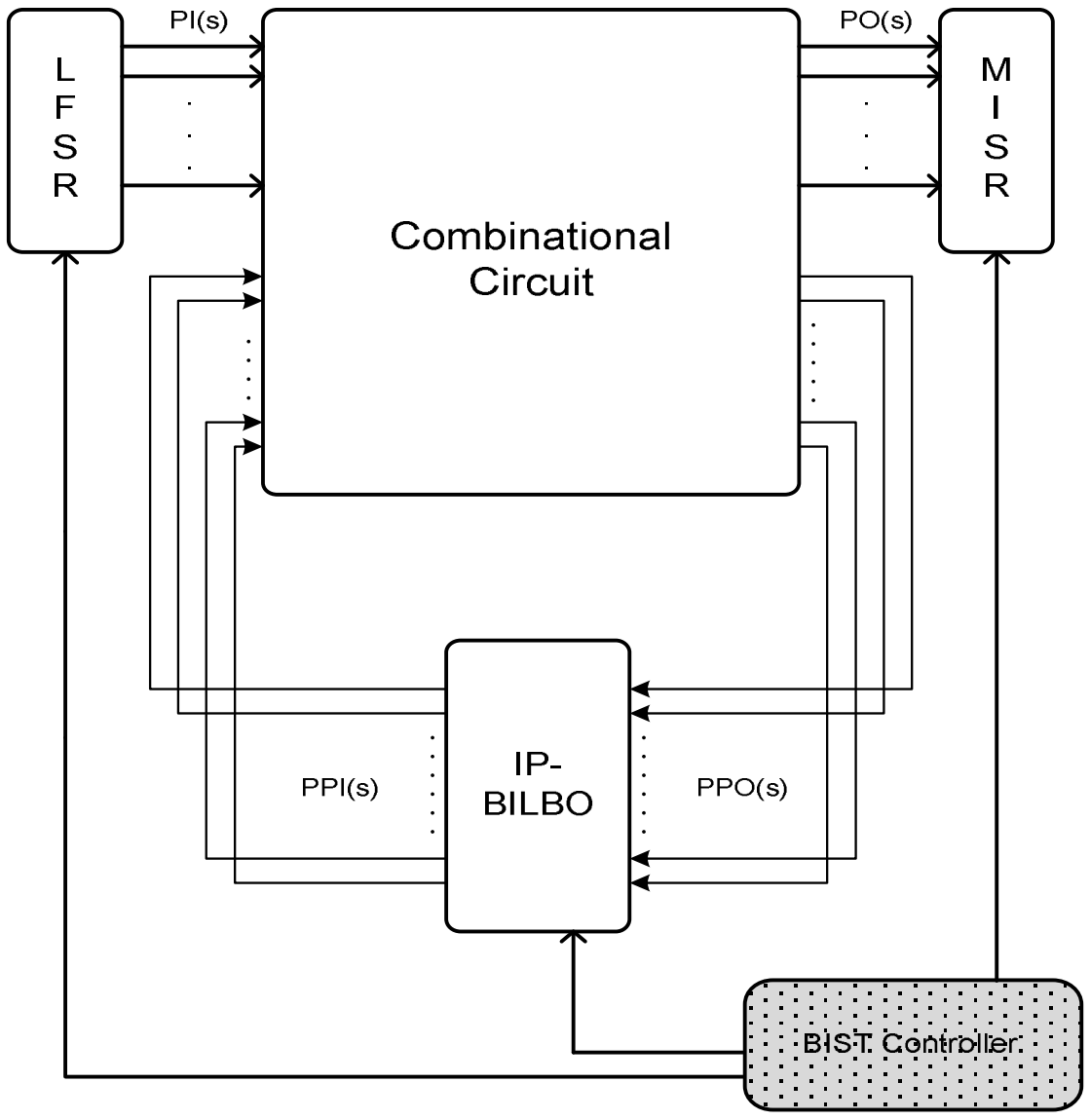

Fig. 1. Architecture of BIST inserted system.

In this figure, the system block diagram is shown in test mode. The system is illustrated in Huffman model which separates the system into combinational and sequential parts. PIs are primary inputs of the circuit that are fed using a LFSR. POs are primary outputs of the circuit that are fed to a MISR, and the BIST controller controls the test process. The sequential part of the system is integrated with a modified version of BILBO that we refer to as IP-BILBO. Fig. 2, depicts the internal structure of a common BILBO.

In a regular BILBO as shown in Fig. 2, $B_2B_1$ = 10 resets the internal registers, 01 configures BILBO as a MISR, 00 enables serial shift-in mode, and 11 configures the BILBO as a register with parallel loading. The latter mode is the normal mode for the operation of the sequential circuit being tested.

For the test mode the proposed method (in its first phase) uses of MISR mode of BILBO as a test pattern generator like a LFSR. This structure generates a pseudo random test vector in two cycles. In the first cycle for $B_2B_1$ = 11, PPOs of the circuit store in the internal registers of the BILBO. In the next cycle for $B_2B_1$ = 01 a value, mixed from feedback circuit and values of PPOs stores in registers. This architecture uses this value as a pseudo random test data. It is worth mentioning that in this structure number of cycles for pseudo random test generation can vary from one cycle to any number of cycles. As the number of cycles to generate a random number increases the achieved pattern would be more randomized.

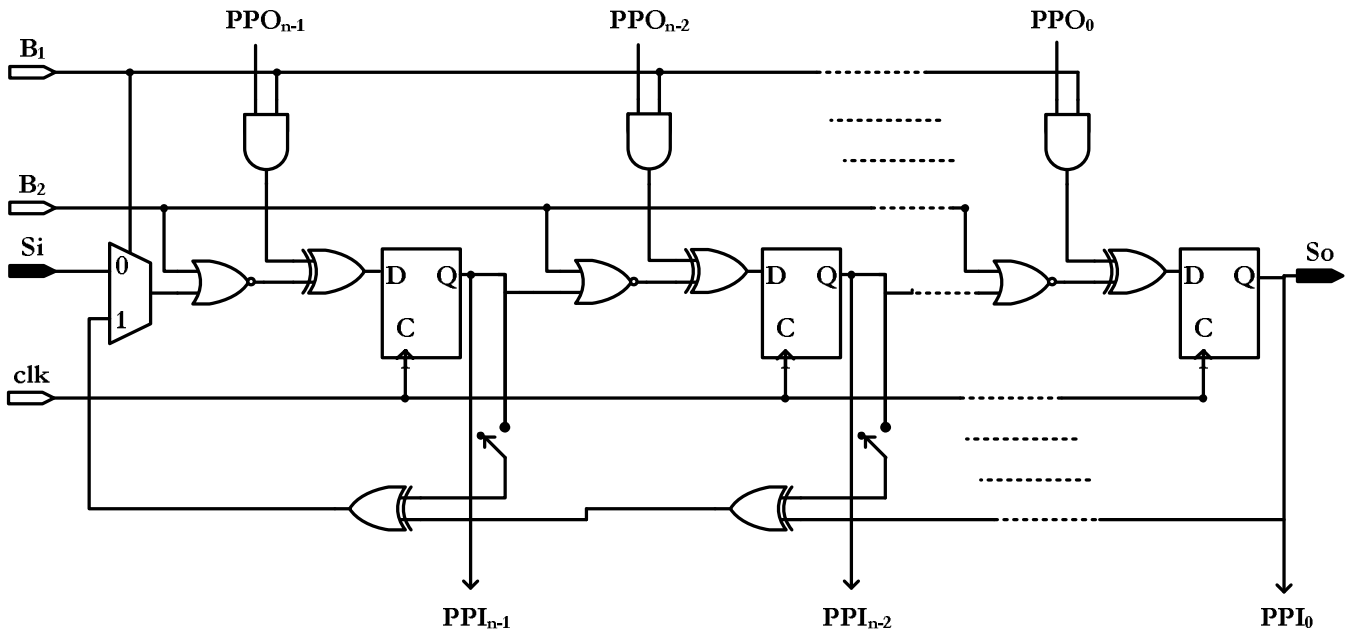

Fig. 2. Schematic of a common BILBO.

Use of this structure as a PRPG, as described earlier, could be problematic. There are scenarios in which the circuit under test enters in an infinite loop state and values on PPIs and PPOs repeat boundlessly. One of these scenarios is shown in Fig. 3. Although in circuits with a large number of internal signals such cases may seem rare, the situations with this property have to be dealt with. In Fig. 3, it is assumed that polynomials are in a way that the BILBO circuit produces 101 as output for 110.

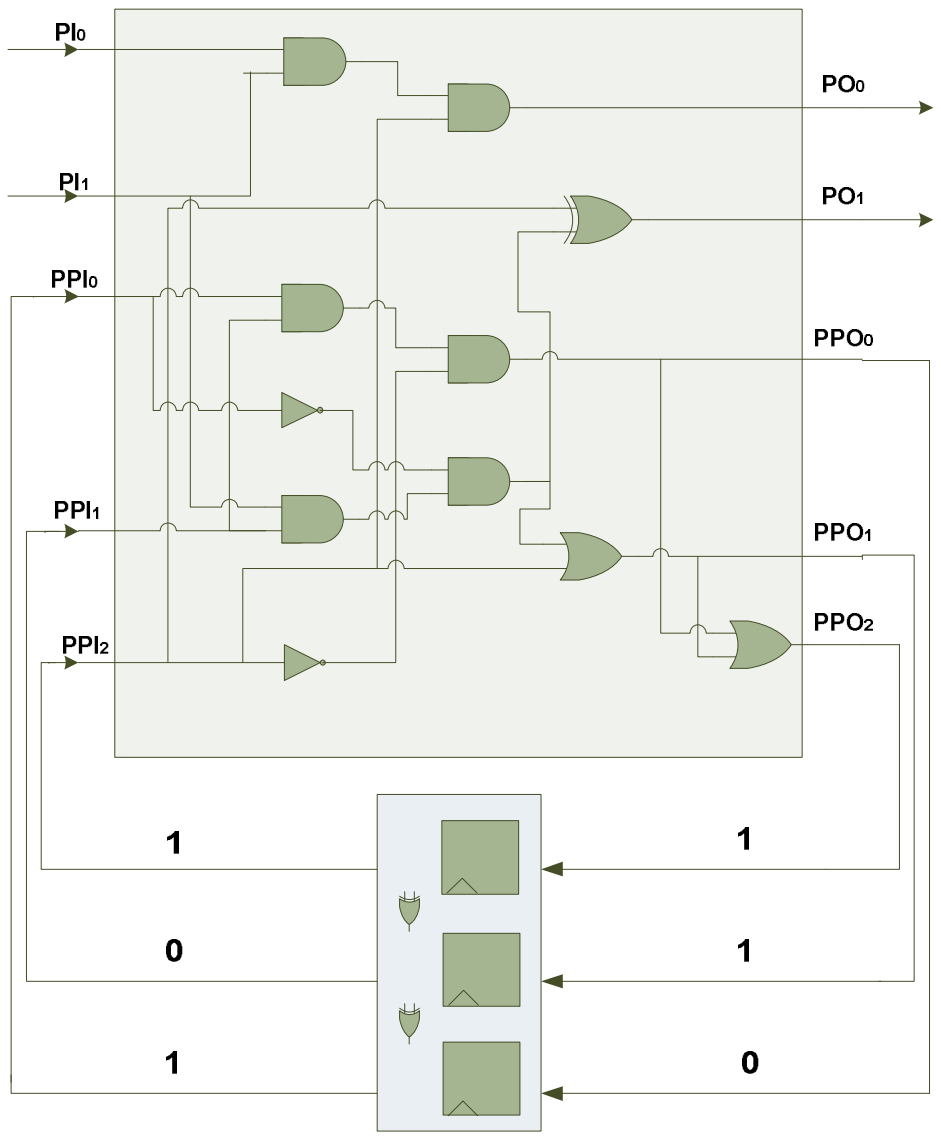

Fig. 3. A sequential circuit with repeating loop problem.

In this figure, when PPIs of the circuit become 101, the result values on PPOs will be 110. This condition can significantly degrade test coverage results.

For solving this problem we have proposed a new structure for BILBO that we refer to as an irregular polynomial BILBO (IP-BILBO). The architecture of IP-BILBO is shown in Fig. 4.

This structure is shown in Fig. 4 benefits from irregular polynomial reseeding. In this structure implemented by the multiplexers, from each pseudo random test data to the next, polynomials can differ. Because this structure does not directly reseed the internal registers of BILBO by outputs of the circuit (PPOs), it prevents it from the infinite loop.

In the proposed IP-BILBO, the modes-select input (MS) = 0, uses value 01 as a pseudo random pattern generator. This mode is the same as the BILBO shown in Fig. 2.

In the second mode, MS = 1, PPOs can directly take part in building of feedback signature. This structure takes only one cycle for generating pseudo random test data. Assuming there is a value on the internal registers when $B_2B_1$ = 00, in each cycle the internal registers will be loaded with a new pseudo random data. In both modes 0 and 1 of this IP-BILBO the history of the produced pseudo random test patterns builds a signature. This signature shifts out from the chained registers in specific time intervals and compares with the signature of a golden circuit. To reduce the probability of signature aliasing, signatures should be load frequently.

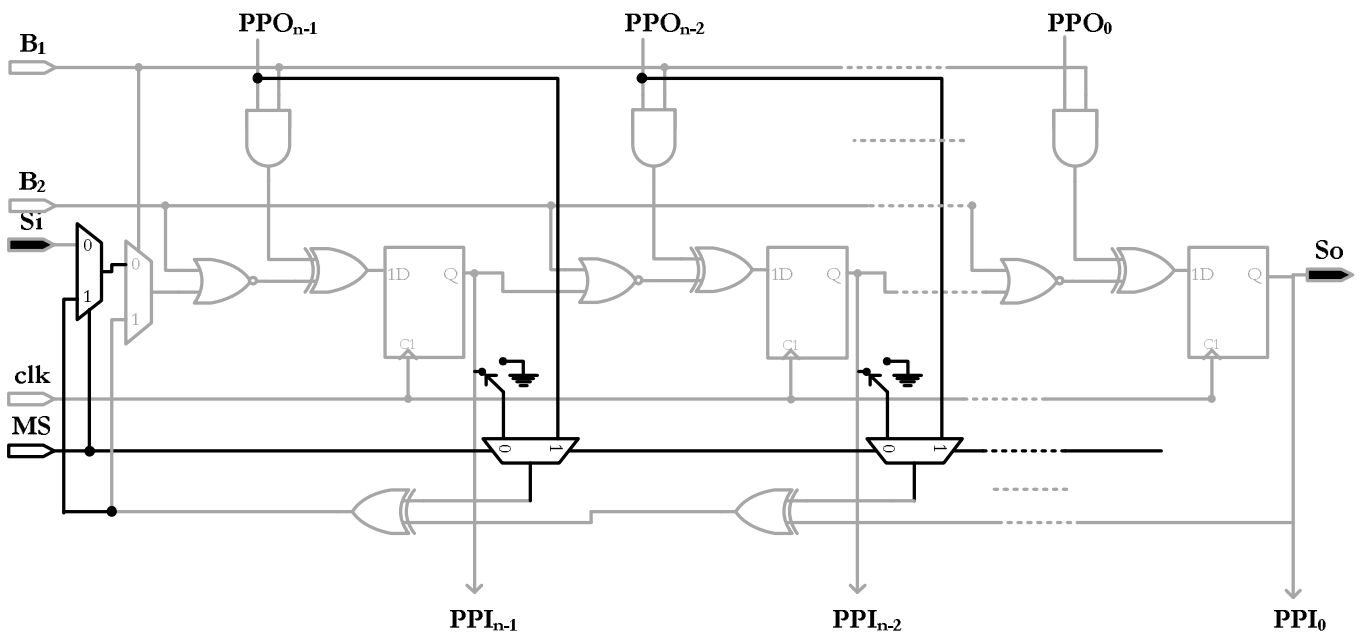

Fig. 4. Schematic of IP-BILBO (added components to BILBO are shown in black).

## III. METHODOLOGY

This section presents a methodology for our proposed test pattern generation and application. The proposed method can be used as a complete hybrid BIST for testing of a sequential circuit.

Fig. 5 depicts an overall view for the testing process. As this figure illustrates the test process consists of two phases. In the first phase, a sequence of deterministic test vectors with pseudo-random test patterns produced by the first mode of IP-BILBO (MS = 0) is applied to the circuit. In the second phase the second mode of IP-BILBO (MS = 1) is selected.

In Phase 1, deterministic test patterns are produced using ATALANTA tool. The test vector which detects most faults in the circuit is selected in the next step. In the third step, this test vector is applied to the CUT and fault coverage is calculated. In the next step, the achieved coverage is compared with the threshold 1 (*th1*) value, and if the required threshold is met the process of testing goes to the second phase, otherwise the process continues to Step 5. In Step 5 the number of continuous pseudo-random produced test patterns which have not detected any faults is compared with threshold 2 (*th2*). The value of *th2* depends on the scan length which is the same as the shifting cycles. In this work, *th2* is chosen as one half of the number of shifting cycles for small circuits and a smaller ratio of the shifting cycles for larger circuits. In Step 5, if the number of iterations becomes larger than *th2*, the circuit uses ATALANTA to generate new deterministic test patterns as before for the remaining. Otherwise, the testing process continues to Step 6 and a pseudo random test data is produced using IP-BILBO in its first mode.

For the second phase, the test application is similar to the first phase except that in this phase the circuit is in the second mode of IP-BILBO. In the second phase, threshold 3 (*th3*) is twice as much as *th2*, because in this phase it takes only one clock cycle to produce a pseudo random test data instead of two cycles in the first phase.

The number of test cycles required in RTS architecture is computed as RTS test cycles (*RTC*) from Equation (1).

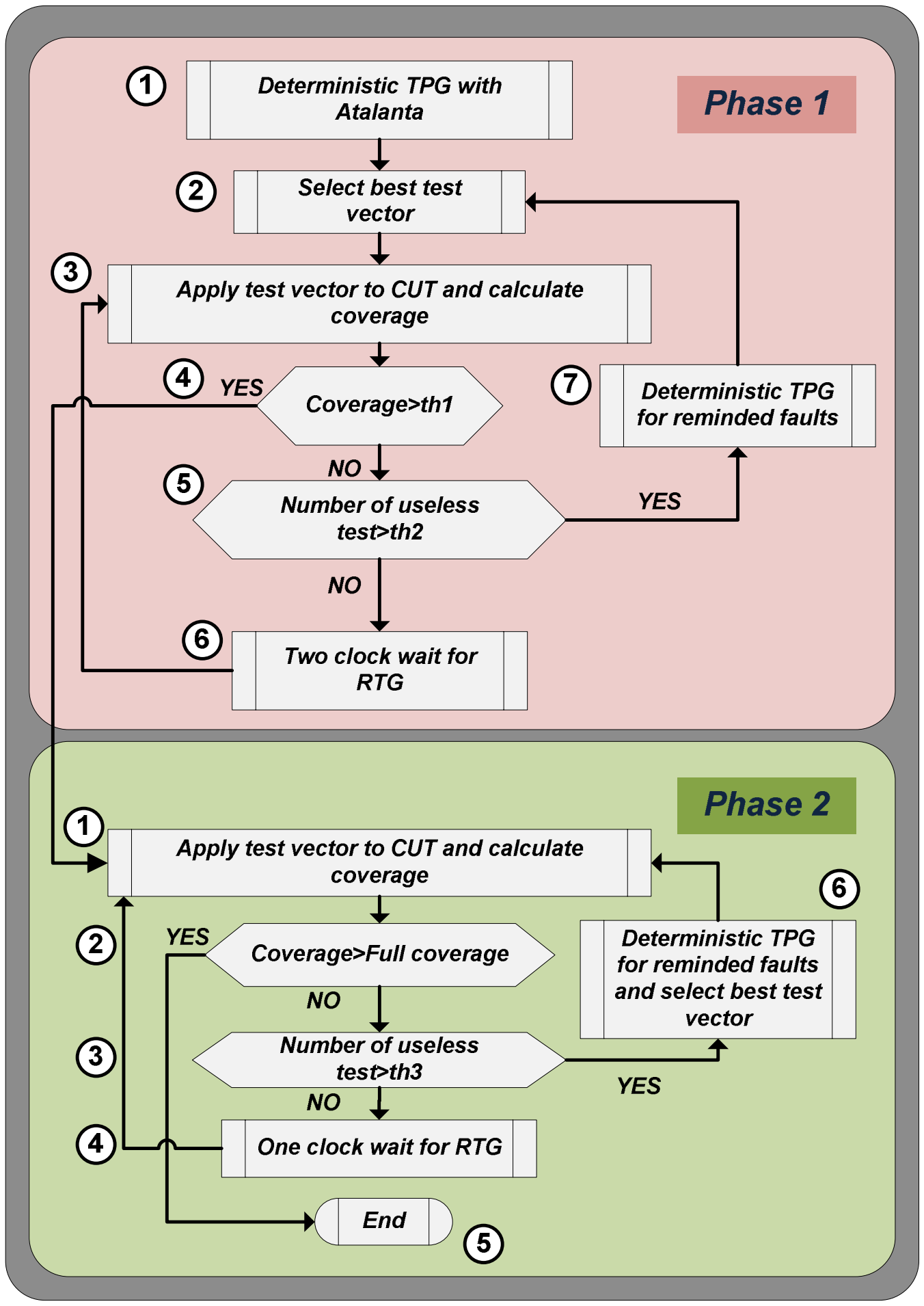

Fig. 5. Test application flowchart

$$RTC = (ADV \times (PIs + PPIs)) \quad (1)$$

In this equation *ADV* denotes to the number of ATLANTA deterministic test vectors. *PIs* are the size of primary inputs of the CUT. *PPIs* are the number of pseudo-primary inputs of the CUT. Size of *PPIs* in a circuit is equal to the number of internal registers.

*PWTC* shows the total number of cycles from a previous work [19] that is computed from Equation (2).

$$PWTC = [(ScanVec \times (PIs + PPIs)) + (CircResp)] \quad (2)$$

In this equation *ScanVec* is the number of selected deterministic test vectors in [19] and *CircResp* is the number of cycles that circuit uses its *PPOs* as *PPIs*.

Finally our proposed method test cycles is computed from Equation (3).

$$PMTC = [(PMDV \times (PIs + PPIs)) + (PRTP_Ph1 \times 2) + (PRTP_Ph2)] \quad (3)$$

In this equation, *PMDV* is the total number of selected deterministic test vectors. *PRTP_Ph1* is the number if pseudo random test vectors generated in Phase 1 and *PRTP_Ph2* is the number of pseudo random test vectors generated in Phase 2. The percentage of improvement of the work presented in this paper in proportion to RTS is computed from Equation (4).

$$RTS\,Imp = \frac{(RTC - PMTC)}{RTC} \times 100 \quad (4)$$

And Equation (5) shows the improvement in proportion to work presented in [19].

$$PW\,Imp = \frac{(PWTC - PMTK)}{PWTC} \times 100 \quad (5)$$

## IV. Experimental Results

The experimental results for our hybrid BIST method for ISCAS'89 benchmark circuits are presented in this section. Fig. 6, shows the amount of reduction in test time in our work and the work in [19] in proportion to RTS.

In this figure the gray bars show the percentage of test time reduction in previous work in relation to RTS method, and the hatched bars show the percentage of test time reduction in our method. More details about results can be inferred from TABLE I. The fault coverage for the benchmark circuits has been considered full.

## V. Summary and Conclusion

As the technology advances, testing of sequential circuits has remained a challenging problem. Also systems tend to become larger and more complex in each technology generation. This situation has made BIST structures as a crucial system component. The proposed IP-BILBO in this work attempts to mitigate this issue by local test data generation. This method makes use of two techniques of direct and indirect reseeding of internal registers for producing pseudo random test patterns. This reduces the number of required deterministic test data significantly. Because of reduction in number of deterministic test data the cost of communication decreases significantly specially in large systems like SoCs and MPSoCs.

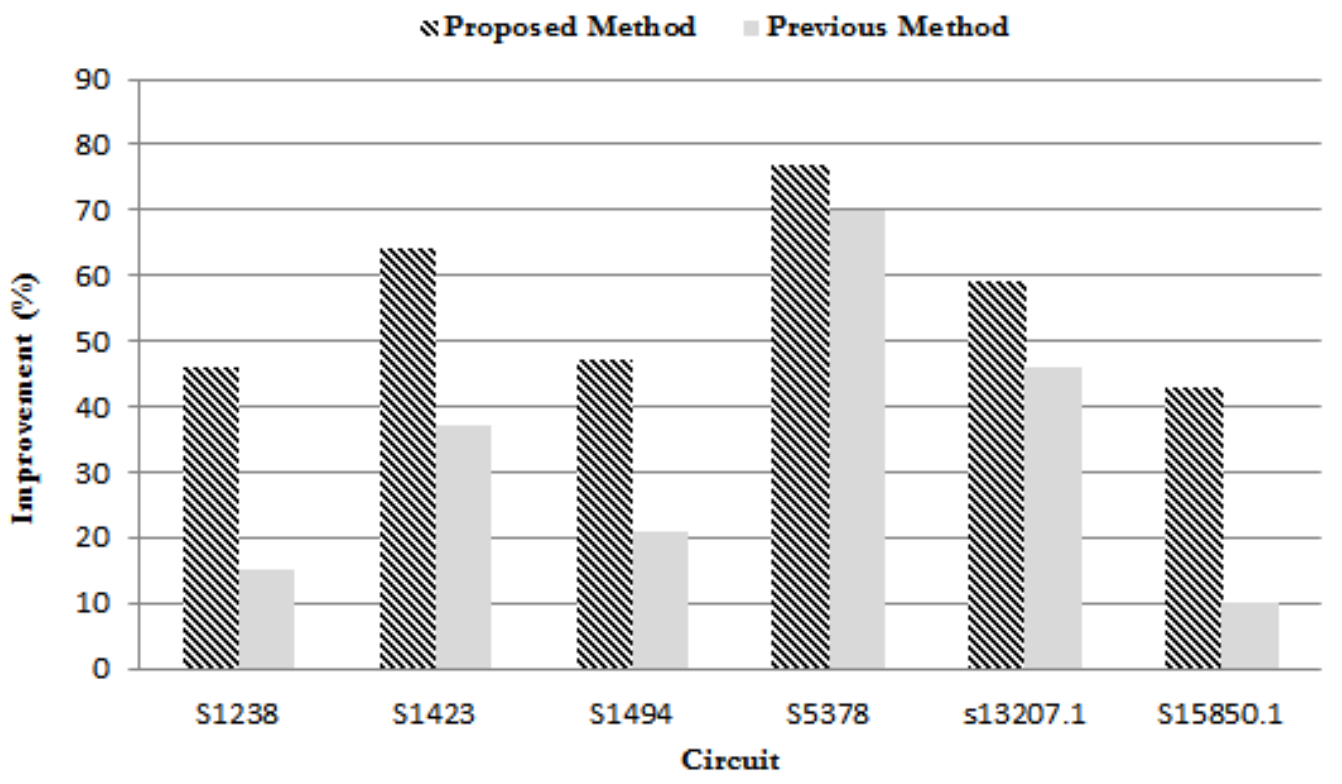

Fig. 6. Percentage of improvement in proportion to RTS.

TABLE I. Test application time improvement for proposed method

| Circuit | #ADV | #PRTP | #PMDV | #RTC | #PWTC | #PMTC | Imp to RTS (%) | Imp to [19] (%) | PIs+PPIs | #Faults |
|---|---|---|---|---|---|---|---|---|---|---|
| S1238 | 149 | 726 | 58 | 4768 | 3052 | 2583 | 46 | 15 | 32 | 1355 |
| S1423 | 69 | 916 | 13 | 6279 | 3543 | 2233 | 64 | 37 | 91 | 1515 |
| S1494 | 129 | 275 | 48 | 1806 | 1224 | 966 | 47 | 21 | 14 | 1506 |
| S5378 | 263 | 5270 | 37 | 56282 | 44195 | 13189 | 77 | 70 | 214 | 4551 |
| s13207.1 | 466 | 4740 | 185 | 326200 | 246183 | 134241 | 59 | 46 | 700 | 9815 |
| S15850.1 | 448 | 2696 | 249 | 273728 | 172349 | 154840 | 43 | 10 | 611 | 11725 |

- ADV: The number of deterministic test vectors generated by ATALANTA.
- PRTP: The number of pseudo random test vectors in phase1 and phase2 of proposed method.
- PMDV: The number of deterministic test vectors that generated by proposed method.
- RTC: RTS Test Clock.
- PWTC: Previous Work Test Clock.
- PMTC: Proposed Method Test Clock.
- Fault coverage has been considered full.